\documentclass[twocolumn,aip,cha,amsmath,amssymb,reprint]{revtex4-1}
\usepackage{graphicx}
\usepackage{dcolumn}
\usepackage{bm}
\usepackage{color}

\usepackage{pseudocode}

\usepackage{changepage}

\usepackage{cases}

\begin{document}

\title{A unified method of detecting core-periphery structure and community structure in networks}

\author{Bing-Bing Xiang}

\affiliation{School of Mathematical Science, Anhui University, Hefei
230601, P. R. China}

\author{Zhong-Kui Bao}

\affiliation{School of Mathematical Science, Anhui University, Hefei
230601, P. R. China}

\author{Chuang Ma}

\affiliation{School of Mathematical Science, Anhui University, Hefei
230601, P. R. China}

\author{Xingyi Zhang}

\affiliation{Institute of Bio-inspired Intelligence and Mining Knowledge,
School of Computer Science and Technology, Anhui University, Hefei 230601, China}

\author{Han-Shuang Chen}
\affiliation{School of Physics and Material Science, Anhui University, Hefei 230601, China}
\author{Hai-Feng Zhang}
\email{haifengzhang1978@gmail.com}
\affiliation{School of Mathematical Science, Anhui University, Hefei
230601, P. R. China}
\affiliation{Center of Information Support \&Assurance Technology, Anhui University, Hefei  230601,  China
}
\affiliation{Department of Communication Engineering, North
University of China, Taiyuan, Shan'xi 030051,  China}

\date{\today}

\begin{abstract}
Core-periphery structure and community structure are two typical meso-scale structures in complex networks. Though the community detection has been extensively investigated from different perspectives, the definition and the detection of core-periphery structure have not received much attention. Furthermore, the detection problems of the core-periphery and community structure were separately investigated. In this paper, we develop a unified framework to simultaneously detect core-periphery structure and community structure in complex networks. Moreover, there are several extra advantages of our algorithm: our method can detect not only single but also multiple pairs of core-periphery structures; the overlapping nodes belonging to different communities can be identified; different scales of core-periphery structures can be detected by adjusting the size of core. The good performance of the method has been validated on synthetic and real complex networks. So we provide a basic framework to detect the two typical meso-scale structures: core-periphery structure and community structure.
\end{abstract}


\maketitle

\begin{quotation}
Community structure in complex networks as a typical meso-scale structure has received considerable attention, however, the other type of meso-scale structure---core-periphery  (abbreviated as CP) structure has received much less attention than they deserve, even though CP structure has been extensively observed in many social and biological systems. Recently, some researchers have begun to pay attention to this problem and designed some algorithms to detect CP structure. Nevertheless, most of them aimed at detecting single CP structure but cannot detect multiple pairs of CP structures. In addition, the two meso-scale structures are separately studied previously. In this work, we propose a simple and effective algorithm which can not only detect single as well as multiple pairs of CP structures, but also can simultaneously detect CP structure and community structure in complex networks. Moreover, the multi-scale of CP structure and the overlapping nodes can be detected too. Therefore, our proposed algorithm builds a bridge connecting the two typical meso-scale structures: CP structure and community structure.
\end{quotation}

\section{Introduction}
Many real-world systems in the field of communication, social, transportation, information, biology and so on, can be described as networks~\cite{newman2010networks}. Meso-scale structures are very important for understanding of network properties and dynamics. Considerable investigations have focused on the study of a particular type of meso-scale structure known as community structure which has some cohesive groups called ``communities'', nodes in the same community are connected densely to each other, whereas nodes in different communities are connected sparsely~\cite{girvan2002community,newman2004finding}. The adjacency matrix representing community structure is given in Fig.~1(a). Various techniques have been developed to detect community structure~\cite{fortunato2010community,harenberg2014community}, e.g., the algorithms were based on modularity~\cite{newman2006modularity}, random walks~\cite{rosvall2008maps}, spectral clustering~\cite{yang2012spectral}, hierarchical clustering~\cite{lancichinetti2009detecting,yang2013hierarchical}, nonnegative matrix factorization approach~\cite{yang2013overlapping} and so on. There are also many algorithms for overlapping community detection based on label propagation~\cite{gregory2010finding,he2014node}, link partition~\cite{ahn2010link,yu2013topic}, clique percolation theory~\cite{palla2005uncovering,zhang2017fast}, multi-objective evolutionary algorithm~\cite{zhang2017mixed},  etc.

Another meso-scale structure known as core-periphery  structure has not received enough attention, but it has been examined in the networks of society~\cite{white1976social}, scientific citation~\cite{doreian1985structural}, international trade~\cite{nemeth1985international,smith1992structure} and other fields~\cite{doolittle1996self,barucca2016disentangling,holme2005core}. For instance, maintaining a party or an organization often needs an elite group to organize and manage it. The elite group is composed of highly influential and powerful individuals who have more connections with each other. And other people involved, namely the periphery part with few connections inside are less organized and less dominant~\cite{smith1992structure}. Biological networks, such as that of the human brain, where a group of densely connected network nodes provides long-term functionality and robustness (core), while another group of sparsely connected nodes  is responsible for adaptation on short time scales to changing conditions in the environment (periphery)~\cite{otokura2016evolutionary}. All of these phenomena are the expressions of CP structure, indicating the significance of studying the CP structure~\cite{borgatti2000models,rombach2014core,csermely2013structure}.

Although the two meso-scale structures were studied more or less, the relationship between them is not well clarified. In fact, there are some intuitive relationships between the two typical meso-scale structures: on the one hand, a network with a single CP structure (the adjacency matrix is shown in Fig.~1(b)) implies the absence of community structure; on the other hand, a network with multiple pairs of CP structures (the adjacency matrix is illustrated in Fig.~1(c)) often implies the existence of community structure. However, community structure does not imply the existence of multiple CP structure, if there are no closely connected core nodes and more loosely connected peripheral nodes in each community, the multiple pairs of CP structure does not appear. Therefore, correct understanding the relationship between two meso-scale structures can help us distinguish and design effective methods to detect both of them.

\begin{figure}
\centerline{\includegraphics[width=3.5in]{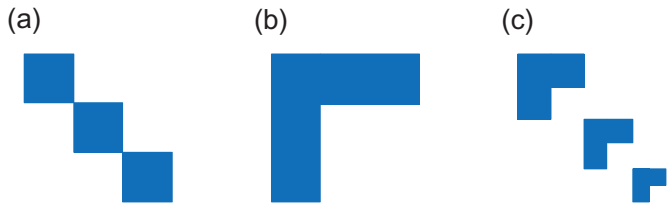}}
\caption{Adjacency matrices of idealized block models. (a) community structure, (b) single CP structure, (c) multiple pairs of CP structure.}
\label{fig1}
\end{figure}
The formal and rather popular definition of CP structure was proposed by Borgatti and Everett~\cite{borgatti2000models}, in which a node
belongs to a core if and only if it is well connected both to other core nodes and to
peripheral nodes, and peripheral nodes do not connect with other peripheral nodes. The adjacency matrix of an ideal CP structure is illustrated in Fig.~1(b)~\cite{rombach2014core}. Since the definition of ideal CP structure is too strict to meet, a looser definition of CP structure is that core nodes are highly interconnected and peripheral nodes frays into a tree~\cite{verma2016emergence}. However, it is a descriptive definition, the strict definition from mathematics has not been well proposed. The existing CP detection methods were mainly implemented by checking how well a network approximates the ideal case~\cite{borgatti2000models,rombach2014core,della2013profiling} or by ranking the nodes according to certain centrality indices denoting the nodes' coreness~\cite{da2008centrality,della2013profiling,cucuringu2014detection}. Recently, Zhang \emph{et al.} first proposed a method for identifying the CP structure using a maximum likelihood method to fit a stochastic block model with CP structure. By fitting
the model to the observed network data, the parameters of the fit
tell us the best partition with respect to the CP structure~\cite{zhang2015identification}. The method is nonparametric, and can deal with large-scale networks and weak CP structure. However, these methods often face one or several of following shortcomings: first, the methods need to give the size of core in advance; second, the methods rudely divide a network into a single CP structure, however, many real networks present multiple pairs of CP structures~\cite{kojaku2017finding}. More importantly, multiple pairs of CP structures and community structure may coexist in many real networks, the methods do not provide a unified framework to detect the both meso-scale structures.

Inspired by these reasons, the goal of this paper aims to propose a unified method to detect CP structure and community structure. In doing so, we firstly rank all nodes according to a connection density indictor, then we can judge whether the network exhibits a single CP structure, multiple pairs of CP structures or community structure based on the defined region density curve. Moreover, the multi-scale of CP structure and the overlapping nodes can be detected too.

\section{The Proposed Algorithm}\label{sec:alg}

In this section we describe the method used for detecting both CP structure and community structure. Consider an undirected and unweighted network, the general framework of our method is presented in Algorithm ~\ref{algorithm1} (shown in the Appendix). The algorithm consists of three main steps: 1) re-rank all nodes in a new sequence; 2) the region density of each node is calculated, and the region density curve is plotted; 3) detect main meso-scale structures of the network based on the region density curve.

\subsection{Re-rank nodes in a new sequence}\label{sec:rerank}
Consider a network $G(V,E)$, where $V$ is the set of nodes and $E$ is the set of links. Given that the links between core nodes are denser, so we want to re-rank all nodes such that nodes with more common connections approach each other in the new sequence. In doing so, we define a set $U$ and a set $V'=V\setminus U$ to store re-ranked nodes and the remaining nodes, respectively. Initially, $U=\emptyset$ and $V'=V$. To start the sorting process, we need to choose one node as the starting node. We can choose a node with the highest centrality value since the node is more likely to be core node.
We here choose the node with the highest closeness centrality as the first node in set $U$, and renumber it as $u_1$, i.e., $U=\{u_1\}$ (we found that the results are not sensitively dependent on different centrality indices, such as degree, betweenness, eigenvector, and so on~\cite{newman2010networks}). Now we need to choose a node from $V'$ and put it into $U$ to make sure the new added node has the most connections with the nodes in $U$. If more than one nodes are found, we choose the node with the maximum degree and put it into $U$.  Namely, choose the node with the largest value of $P$ and put it into set $U$, which is defined as:

\begin{equation}\label{eq1}
\centering
P\left({i}\right)=\sum\limits_{j\in U}A_{ij}+k\left({i}\right)/k_{max},
\end{equation}
where $A_{ij}$ is the element of adjacency matrix, and $k(i)$ and $k_{max}$ are the degree of node $i$ and the maximal degree, respectively.

The framework of re-ranking process is given in Algorithm~\ref{algorithm2} (shown in the Appendix). Take an illustration in Fig.~2 as an example, there are four red nodes in $U$, now one node in $V'$ (outside of the circle) is going to be added into $U$ (see Fig.~2(a)). However, nodes $a$ and $b$ have the same connections with $U$, we choose node $a$ and put it into $U$ since node $a$ has larger degree value (see Fig.~2(b)). If their degree values are also the same, one of them is randomly chosen. In this way the nodes who have more common connections are able to close each other in the new sequence. According to the above method, all nodes are sorted in $U$ as: $U = \left\{ {{u_1},{u_2},\cdots,{u_N}} \right\}$.
\begin{figure}
\centerline{\includegraphics[width=3.5in]{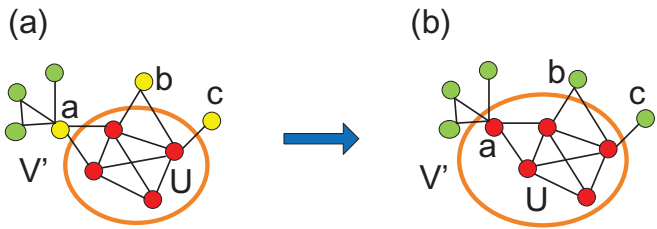}}
\caption{An illustration of re-ranking nodes in a new sequence. (a) red nodes in circle are in set $U$, one node connecting set $U$ is going to be added into $U$, i.e., node $a$, $b$, and $c$, (b) node $a$ is added into $U$ since whose degree is larger than node $b$ even though both of them have 2 connections with $U$.}
\label{fig2}
\end{figure}

\subsection{Plot region density curve}\label{sec:plot}
We here first define a local indicator---connection density (CD) to characterize the density of connections in a subgraph $S$, which is given as:
\begin{equation}\label{eq2}
\centering
CD\left( {S} \right) = \frac{{2m'}}{{n'\left( {n' - 1} \right)}},
\end{equation}
where $n'$ is the number of nodes in $S$ and $m'$ is the number of existing connections in $S$.

A parameter $\alpha$ is defined to measure the minimal size of core. For a given value of $\alpha$, the region density of a node $u_i$ is defined as the connection density of subgraph consisting of nodes from $u_{i-\alpha+1}$ to itself, i.e.,
\begin{equation}\label{eq3}
RD \left(u_i\right)= \left\{ \begin{array}{l}
CD\left( {\left\{ {{u_1}, \cdots ,{u_i}} \right\}} \right),\quad \quad\quad i\le\alpha;\\
CD\left( {\left\{ {{u_{i - \alpha + 1}}, \cdots ,{u_i}} \right\}} \right),\quad i >\alpha.
\end{array} \right.
\end{equation}

We mainly set $\alpha=\lfloor \langle k\rangle\rfloor$ in this work, with $\langle k\rangle$ be the average degree of network and $\lfloor \cdot\rfloor$ be the integral function (In the next context, we also address that different scales of meso-scale structures can be observed by adjusting the value of $\alpha$). We can draw a region density curve after calculating the value of RD regarding each node, where the horizontal axis denotes the node sequence defined in Sec.~\ref{sec:rerank}, and the ordinate axis is the value of RD.

We take the karate club network as an example to illustrate the main steps in the above two subsections. The karate club network consists of 34 nodes that represent club members and 78 links that represent friendships among members. The club was split into two groups because of a conflict of the club president (node 34) and the instructor (node 1), as shown in Fig.~3(a). Note that node 9 is a special node, who supported the president (node 34), but joined the instructor's ( node 1) club for some reasons. To fairly compare with existing methods, we also assume  that node 9 belongs to the node 34's club~\cite{zachary1977information,peel2017ground,girvan2002community}. According to our defined re-ranking way in Sec.~\ref{sec:rerank}, node 1 is firstly put into the set $U$ owing to its the highest closeness value (i.e., $u_1=1$). Though there are a lot of nodes in $V'$ connecting to node 1, node 3 is secondly added into $U$ since it has the maximal degree value (i.e., $u_2=3$). At this time, nodes 1 and 3 are included in set $U$. Next, nodes 2, 4, 14, 8 and 9 in $V'$ connecting to all of nodes in $U$, however, node $2$ is added into $U$ since its degree value is the largest. Repeat the above steps until $V'=\emptyset$. According to the descriptions in Sec.~\ref{sec:plot}, we draw a region density curve in which the ordinate labels the value of RD regarding each node (as shown in Fig.~3(b)).

\begin{figure*}
\centerline{\includegraphics[width=6in]{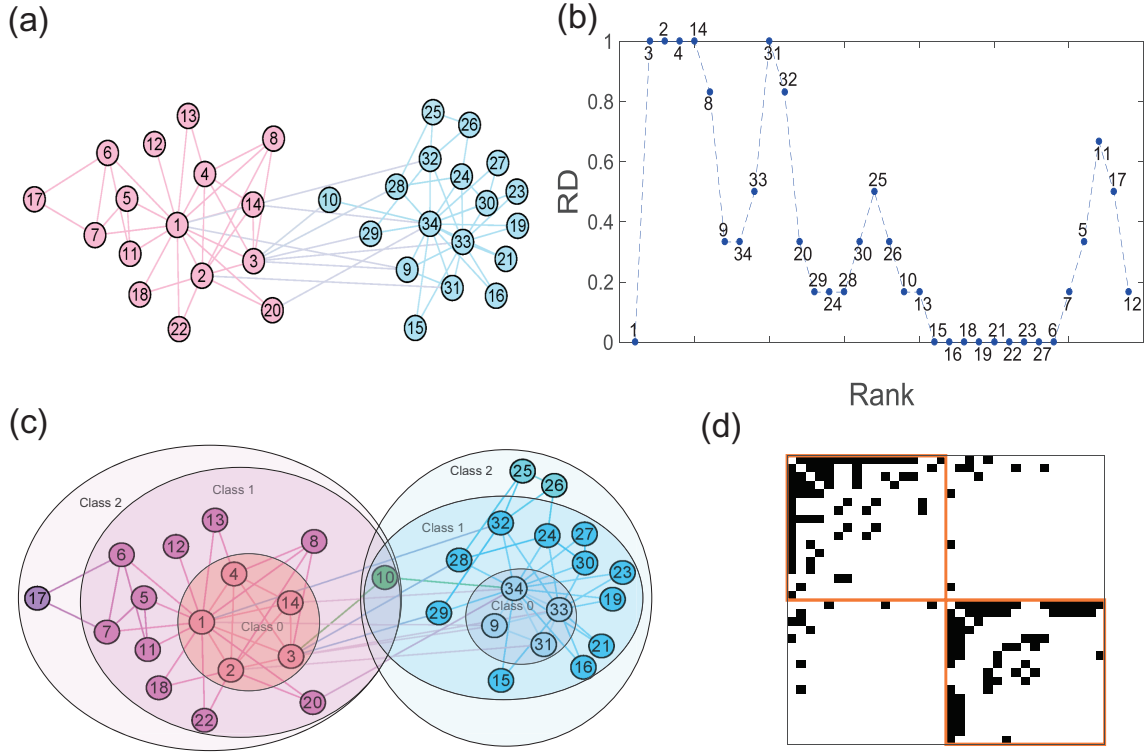}}
\caption{Detection on the karate club network. (a) original structure, where nodes with the same color are in the same community, (b) region density curve of the network, the numbers in the curve are their original numbers, (c) network has two pairs of CP structures, and all nodes are assigned a class value. In addition, node 10 is an overlapping node, (d) matrix representation of the network, where the orange lines indicate the
borders separating different blocks. Here $\alpha=4$ and $\beta=1$.}
\label{fig3}
\end{figure*}

\subsection{Detect meso-structures}
A subgraph $S$ can be viewed as a core if the $CD(S)$ is larger than a threshold value $\beta$. Larger value of $\beta$ gives rise to the stricter definition of core. Once the region density curve is presented, the meso-scale structures can be detected by comparing how many peaks are larger than or equal to $\beta$. Note that some peaks in the beginning of the curve are not valid even though whose $RD(u_i)\geq \beta$. For example, if we set $\alpha=4$, the first three nodes in the region density curve can not form core even though $RD(u_2)=1$ or $RD(u_3)=1$, since the number of nodes is smaller than the minimal size of core $\alpha$. Moreover, if the values of RD for two sequential nodes are greater than or equal to $\beta$, the two cores or communities are merged as a single one. A region density curve of networks with CP structure has the following characteristic: the difference between maximum and minimum values is very huge, and the values of $RD$ for most nodes are very low. This is consistent with the fact that the number of peripheral nodes is far larger than the number of core nodes, and the connections among core nodes are very dense but among peripheral nodes are very few. The framework of detecting core nodes is summarized in Algorithm ~\ref{algorithm3}( shown in the Appendix).

Fig.~4 schematically shows the region density curves of three types of ideal meso-scale structures.  The region density curves in Fig.~4(a) and Fig.~4(b) correspond to a network with a single CP structure and with multiple pair of CP structures, respectively. The region density curve of a community structure without CP structure is similar to a cosine curve, where the difference between peak and least values is not so huge, as shown in Fig.~4(c). Of course, the number of community structures in network is determined by the number of the peaks which reach the value of $\beta$. In sum, we can judge the meso-scale structures of a network according to its region density curve.
\begin{figure}
\centerline{\includegraphics[width=3.5in]{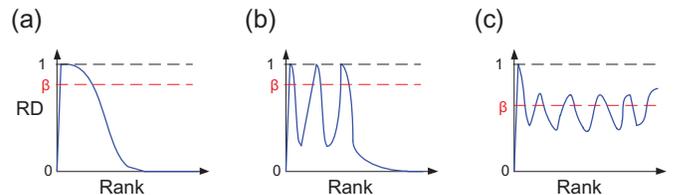}}
\caption{Region density curves of three types of ideal meso-scale structures. (a) single CP structure, (b) multiple sub-CP structures, (c) community structure. The dashed red lines are the threshold value $\beta$, which are used to determine the number of CP structure pairs or communities. }
\label{fig4}
\end{figure}

Though we can determine the core node and the number of cores based on the region density curve, some important problems should be further considered. Firstly, when multiple pairs of CP structures are detected, we should know that the peripheral nodes belong to which core and how close with their core nodes; Secondly, when community structure exists in a network, can we find the overlapping nodes who belong to different communities? To this end, we begin to expand peripheral nodes from each core to form its sub-CP structure. At the beginning, each sub-CP structure only contains core nodes themselves and they are defined as class 0.  The peripheral nodes who have direct connections with core nodes are valued as class 1, and they are allocated to the core which has the most connections with them. In this way, one or several initial sub-CP structures are formed (each sub-CP structure only contains core nodes and their periphery neighbors). Next, the neighbors of the initial sub-CP structures are defined as class 2, and they are allocated to the sub-CP structure which has the most connections with them. The expanding process finishes until all peripheral nodes are allocated and their class values are determined. Many experimental results have demonstrated that most of real networks are ``small-world'', i.e., their average length of paths are very short~\cite{albert2002statistical}. Thus, the above expanding process can be finished in several steps.

In the above allocation steps, some nodes may be allocated to more than one sub-CP structure or community, namely, they have the same neighbors connecting two or several sub-CP/community structures. These nodes are viewed as active nodes. One may intuitively think that these active nodes are the overlapping nodes because they connect different sub-CP structures or communities. However, we address that many active nodes may not be real overlapping nodes. For example, even though an important person need to frequently connect the leaders in different organizations owing to his special role, but which does not mean that the important person belongs to different organizations. Therefore, we re-allocate each active node to a sub-CP/community who has the most neighbors of the active node. If there are still some active nodes who are assigned to more than one sub-CP structure/community, this kind of active nodes are viewed as the real overlapping nodes. The framework of detecting meso-structure is demonstrated in Algorithm~\ref{alogrithm4} (shown in Appendix).

We still use the karate club network as an example to help us understand our method. In Sec.~\ref{sec:rerank}, all nodes have been re-ranked and the region density curve is plotted in Fig.~3(b). Now we should use the number of peaks in the curve to determine the meso-scale structures. According to the assumption $\alpha=\lfloor \langle k\rangle\rfloor$, one has $\alpha=4$ owing to $\langle k\rangle\approx4.59$. Here we set core density $\beta=1$, that is to say, all core nodes are fully connected. As shown in the region density curve, the values of $RD(4)$, $RD(14)$ and $RD(31)$ are equal to 1. Thus, the region formed by these three nodes are $\{1,3,2, 4\}$, $\{3,2,4,14\}$ and $\{9,34,33,31\}$ (though $RD(2)=1$ and $RD(3)=1$, neither of them is valid peak since their region is too small to form core).  Because node 4 and node 14 are sequently placed in the curve, the two cores are merged into one larger core, i.e., $\{1,3,2,4, 14\}$. Now there are two cores: $C_1=\{1,3,2,4,14\}$ and $C_2=\{9,34,33,31\}$ (Fig.~3(c)). According to the above expanding method, we can know that each peripheral node belongs to which core and how far from its own core.

In the first allocation, nodes 10, 28 and 29 are allocated to the two sub-CP structures, so the three active nodes need to be re-allocated. In the secondary distribution, nodes 28 and 29 are allocated to the right sub-CP structure (see Fig.~3(c)). However, node 10 still has the same number of connections to the two sub-CP structures, which is viewed as the real overlapping node.

Fig.~3(d) shows the adjacency matrix of the karate club network, which is very similar to the ideal sub-CP structures shown in Fig.~1(c). Although the karate club network has been used as a benchmark network for community detection, our result indicates that the karate club network has two pairs of CP structures, leading to the network with two communities. We should address that even though some community detection algorithms can divide the karate club network into two groups, which do not clearly clarify each node being core node or peripheral node, namely the inner structure in each group is not well answered. Our method not only divides the network into two groups, but also accurately differentiate the core nodes and peripheral nodes in each group. In addition, our method can further distinguish the roles of peripheral nodes, and identify the active nodes and overlapping nodes simultaneously. In short, our method provides a more systematic and accurate description of the meso-scale structure and the roles of nodes in networks.

\section{Experimental Results}\label{sec:exp}

In this section, we examine the meso-scale structures in some real-world networks and a synthetic benchmark network used in community network detection.

\subsection{USA airport network}
The USA airport network has 332 nodes representing airports and 2126 unweighted links describing the airlines between airports~\cite{zhou2009predicting}. The average degree of this network is $\langle k\rangle=12.81$, so the value of $\alpha$ is 12. By setting $\beta=1$, as shown in Fig.~5(b), only one peak in the region density curve reaches the value of 1, the values of RD for the other nodes are very low. So this network exhibits a strong single CP structure. There are 22 core nodes (the core nodes are the red nodes in Fig.~5(a), and the core nodes are also labeled in the enlarged region density curve (see Fig.~5(c)).

\begin{figure*}
\centerline{\includegraphics[width=6in]{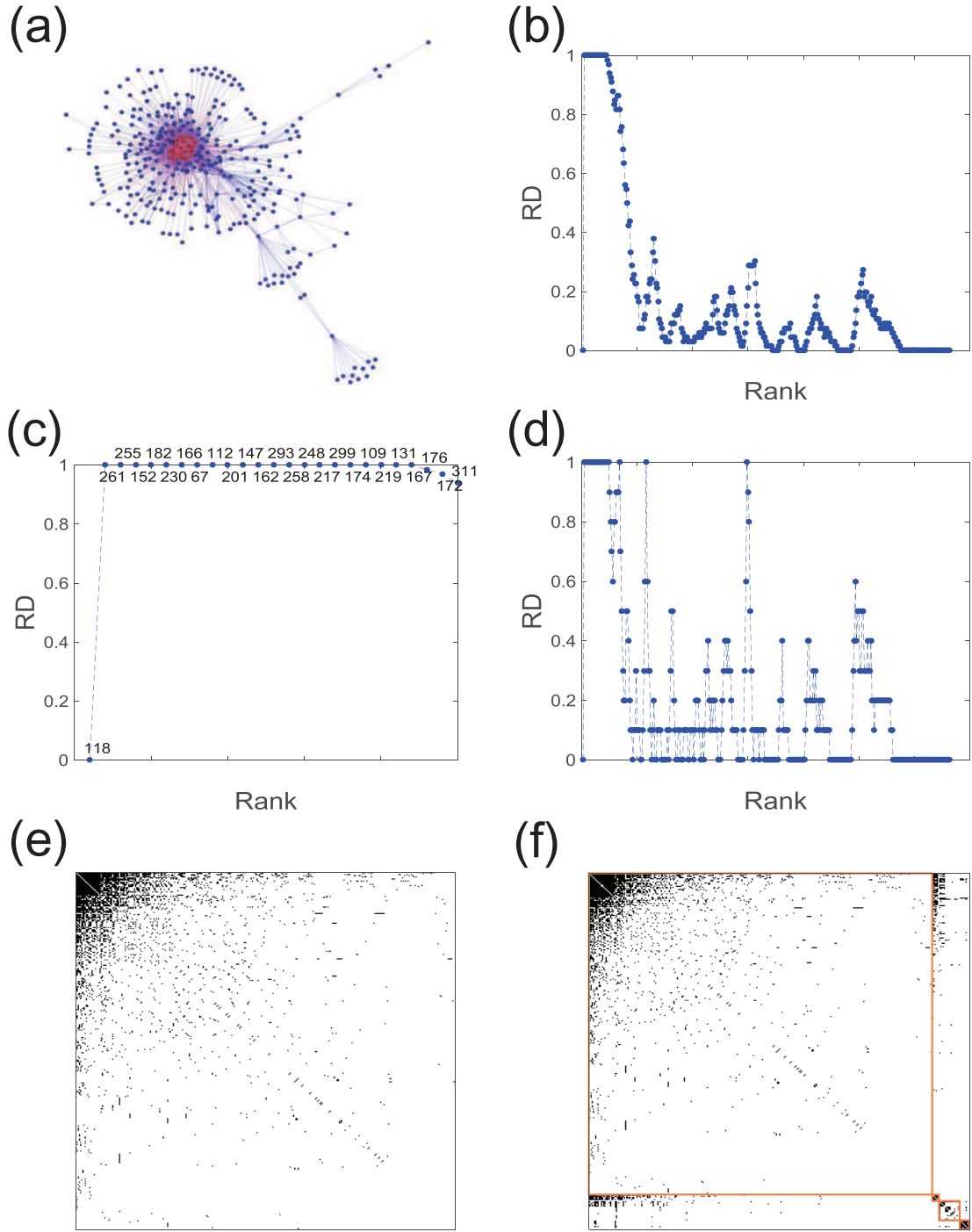}}
\caption{Detection on USA airport network. (a) original structure, where red nodes are core nodes, (b) fully region density curve when $\alpha=12$, (c) region density curve of the first 25 nodes, and the first 22 nodes are core nodes, (d) full region density curve when $\alpha=5$, (e) matrix representation of the network when $\alpha=12$, (f) matrix representation of the network when $\alpha=5$, where the orange lines indicate the
borders separating different blocks. Here $\beta=1$.}
\label{fig5}
\end{figure*}

If we reduce the value of $\alpha$, we can find different sizes of meso-scale structures. When the value of $\alpha$ ranges from 6 to 11, the USA airport network still has a single CP structure, just a slight change of the number of core nodes. When $\alpha=5$, besides the largest CP structure, another three smaller meso-scale structures can be observed. Each of them has 5 core nodes. And their number of peripheral nodes are 3, 14, and 1 respectively. Generally speaking, the number of peripheral nodes is larger than the number of core nodes, so only the second small meso-scale structure can be viewed as a CP structure, the other two are the small scale of community structures. Namely, there are a large CP structure, a small CP structure and two small community structures in this network. Region density curve for $\alpha=5$ is shown in Fig.~5(d), and there are four peaks reaching 1. The first peak contains more nodes, while the other three peaks contain fewer nodes. And the values of RD for most remaining nodes are very low. Figs.~5(e) and (f) show the adjacency matrix of USA airport network for $\alpha=12$ and $\alpha=5$, respectively. The adjacency matrix shown in Fig.~5(e) is very similar to the ideal case shown in Fig.~1(a), indicating this network exhibits a single CP structure. But when $\alpha=5$, there are three smaller meso-scale structures shown in the lower right area of Fig.~5(f).

\subsection{Dolphin social network}

The dolphin social network is an undirected social network of
frequent associations between 62 dolphins in a community
living off Doubtful Sound, New Zealand. The network has 62 nodes representing the dolphins and 159 links denoting the frequent associations between dolphins~\cite{newman2004finding}. The community is composed of two families, as shown in Fig.~6(a), nodes with different colors belong to different families.

The value of $\alpha=5$ in this network. As shown in Fig.~6(b), the difference between the peak and lowest value in region density curve is very huge. The values of RD for most nodes are very low. So we can judge that the dolphin social network has CP structures. If we set $\beta=1$, there are two peaks, but the first peak is invalid since there are only three nodes (nodes 34, 38 and 41). Only the subgraph including nodes 46, 30, 22, 52 and 19 forms a core. So we set $\beta=0.9$ to relax the definition of core. In this case, only one connection is missed in the core group (for a sugraph with five nodes, the possible number of connections is $\binom{5}{2}=10$). Now, $RD(25)=0.9$ and $RD(10)=0.9$, as a result, two cores emerge: $C_1=\{46,30,22,52,19,25\}$ and $C_2=\{14,42,58,18,10\}$, and the network has two pairs of CP structures. Similar to the method in the karate club network, all peripheral nodes can be classified to one corresponding core group and assigned a class value denoting how far from their core group. Fig.~6(d) demonstrates the adjacency matrix of the dolphin social network. If we further relax the definition of core by setting $\beta=0.75$, the network presents a richer phenomenon: there are three pairs of CP structures in the dolphin social network. The accurate partition is given in Fig.~6(e) and its adjacency matrix is presented in Fig.~6(f), respectively. Therefore, more types of partitions can be obtained by adjusting the value of $\beta$.
\begin{figure*}
\centerline{\includegraphics[width=6in]{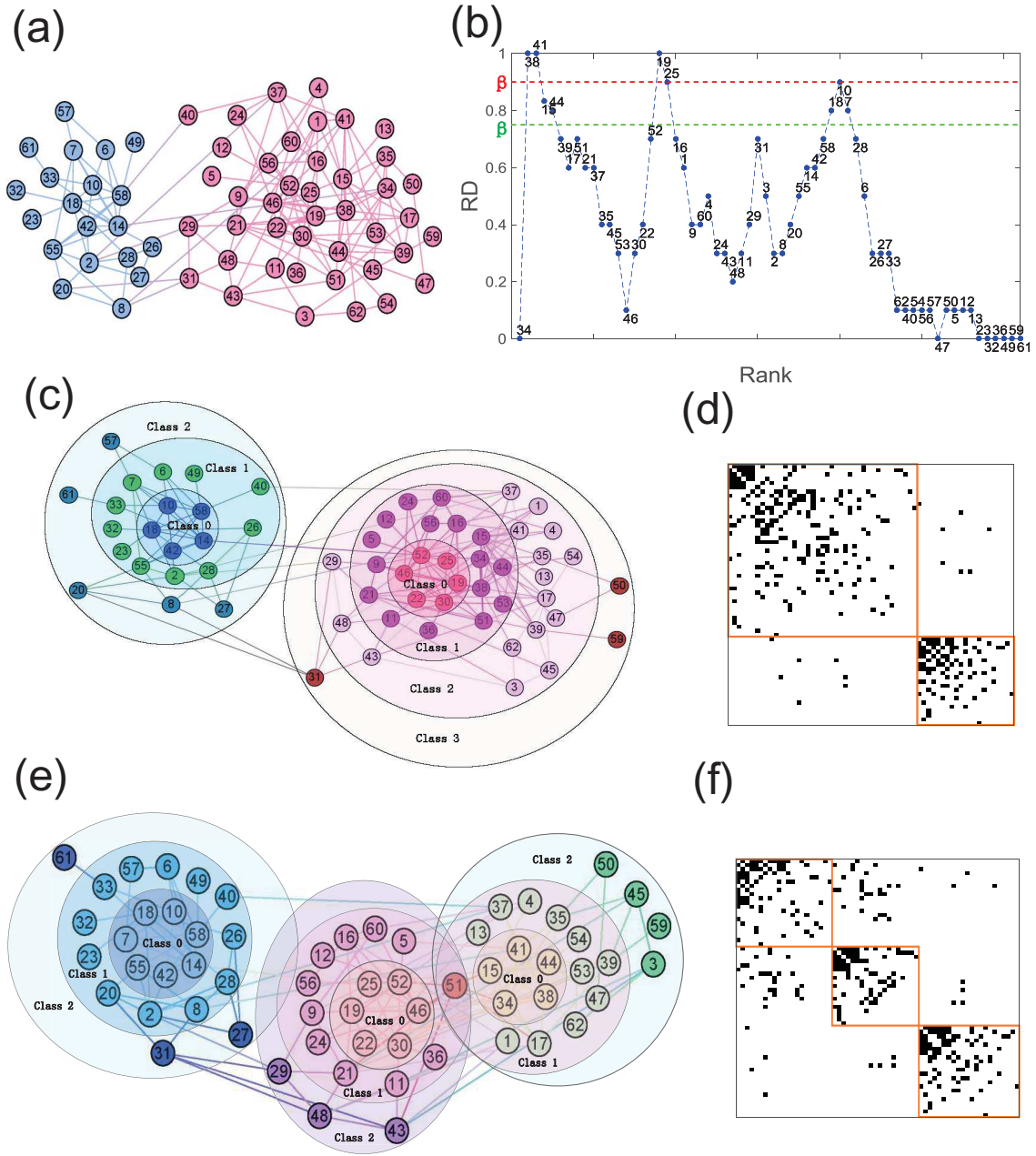}}
\caption{Detection on Dolphin social network. (a) original structure, where the nodes with the same color are in the same community, (b) region density curve, where the dashed red line and the dashed green line correspond to $\beta=0.9$ and $\beta=0.75$, respectively, (c) the network has two pairs of CP structures and each node is assigned a class value,  (d) matrix representation of the network, where the orange lines indicate the
borders separating different blocks, (e) the network has three pair of CP structures and each node is assigned a class value,  (f) matrix representation of the network, where the orange lines indicate the
borders separating different blocks.  $\beta=0.9$ in (c) and (d); $\beta=0.75$ in (e) and (f). Here $\alpha=5$.}
\label{fig6}
\end{figure*}
%
\subsection{Political blogs network}
Adamic and Glance constructed a network of political blogs during
the 2004 U.S. Presidential election. The nodes of this network are blogs about US
politics and the edges are hyperlinks between these blogs~\cite{adamic2005political}. There are 1222 blogs and 16714 connections in the network. Since this network displays a marked division into groups of conservative and liberal blogs, which has been viewed as a typical example of community structure~\cite{zhang2006identification}.

After plotting the region density curve (see Fig.~7(b)), one can see that there are two obvious peaks in the curve. So the network has two pairs of CP structures. By using the expanding method, the peripheral nodes can be divided into their core group and form a corresponding sub-CP structure (see Fig.~7(a)), where blue nodes are the periphery of yellow cores and the red nodes are the periphery of green cores, respectively.  Moreover, there are 11 nodes marked by light blue color are the overlapping nodes. The adjacency matrix in Fig.~7(c) also validates that the network has two pairs of CP structures.

Some methods based on centrality indices assume that the nodes with higher centrality values are core nodes~\cite{da2008centrality}. In this network, we pick two nodes with larger degree values (orange node in left side and the light purple node in right side of Fig.~7(a)) as examples to argue that nodes with larger centrality values are not necessarily the core nodes. The main reason is that these nodes have many connections with peripheral nodes but few connections with core nodes.

\begin{figure*}
\centerline{\includegraphics[width=6in]{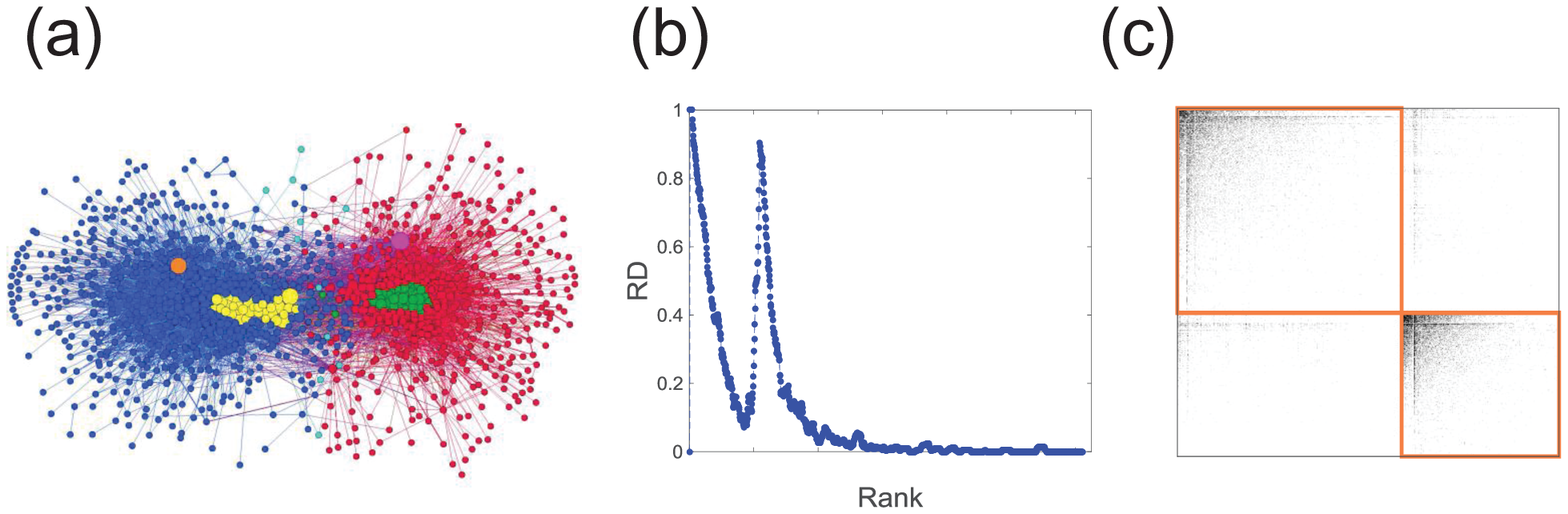}}
\caption{Detection on Political blogs network. (a) Visualization of the network, where yellow nodes and green nodes are two core groups. Blue nodes and red nodes are the peripheries of yellow core and green core, respectively. Light blue nodes are the overlapping nodes. Orange node and light purple node are the nodes with high degree values, their degree values are 301 and 351, respectively,  but they are the peripheral nodes, (b) region density curve of the network, (c) matrix representation of the network,  where the orange lines indicate the
borders separating different blocks. Here $\alpha=27$ and $\beta=0.75$.}
\label{fig_PB}
\end{figure*}

\subsection{Synthetic benchmark network}\label{synthetic}

Our experimental results indicate that our method can not only detect CP structure but also the community structure. But one should note that not all community networks indicate the existence of the CP structure. Here we generate an LFR benchmark network proposed in Ref.~\cite{lancichinetti2009community} to answer the question whether networks with community structure also have CP structure. The LFR benchmark networks is widely used to test the
performances of different algorithms on detecting overlapping communities. In the next section, we also generate several sets of LFR networks to compare the performances of different algorithms on detecting overlapping communities. Here the parameter in this synthetic network are set as follows: the average degree $\langle k\rangle=8$, the maximum degree $k_{max}=12$, the mixing parameter $\mu=0.1$, and the exponents of the power law distribution of node degrees $\gamma_1$ and community size $\gamma_2$ are -2 and -1, respectively. The network has 100 nodes and 404 links, and it is composed of 10 communities whose sizes range from 7 to 12.
The formation mechanism can only guarantee emergence of community structure but cannot ensure there is a core in each community.
The region density curve of the synthetic network is shown in Fig.~8(a), which is very similar to the curve shown in Fig.~4(c). So we can judge the network does not exhibit CP structure. The adjacency matrix of the network shown in Fig.~8(b) is very similar to the Fig.~1(a), which further validates that the network has no CP structure.
\begin{figure}
\centerline{\includegraphics[width=3.5in]{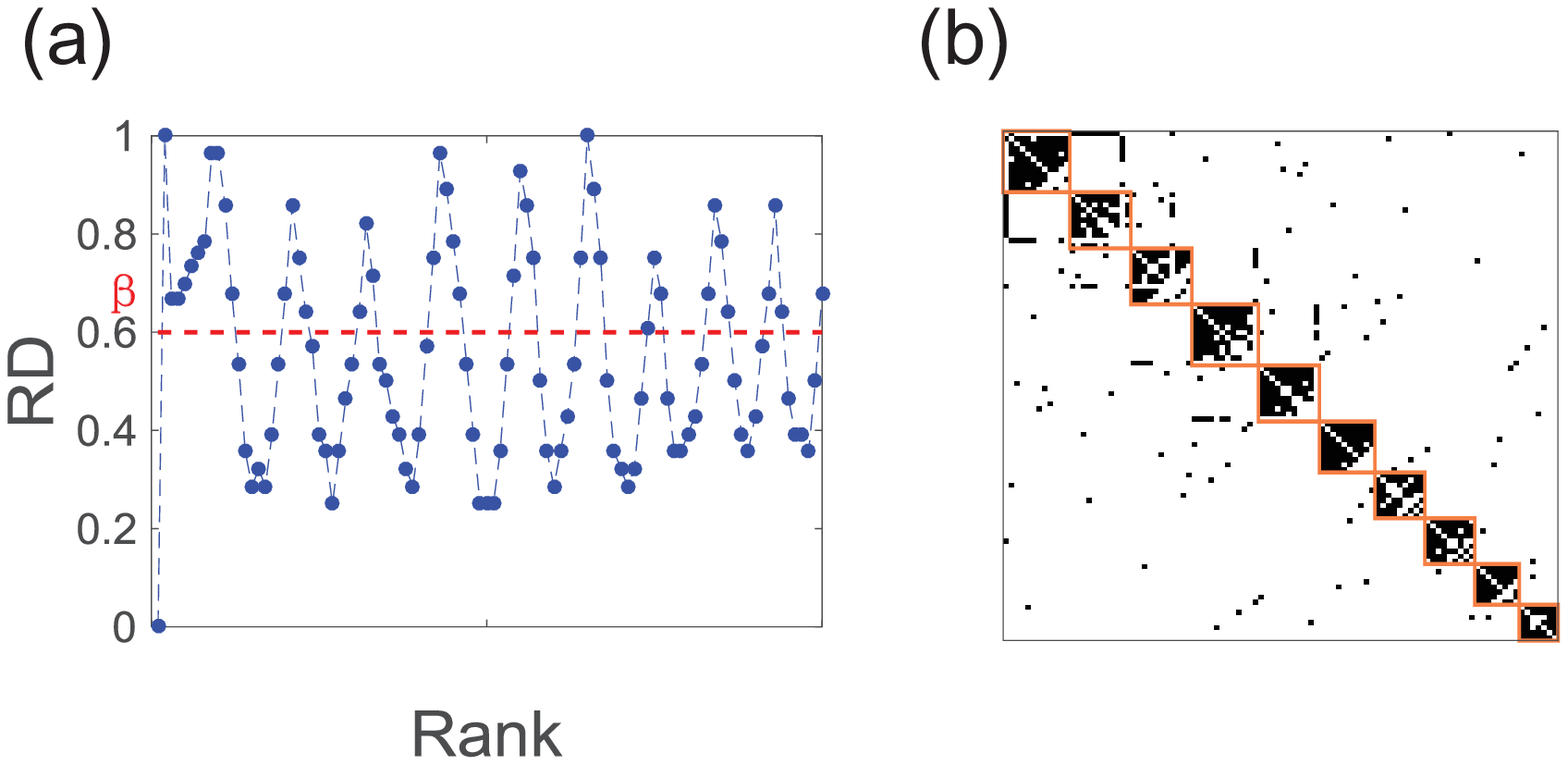}}
\caption{Detection on synthetic benchmark network with  10 communities. (a) region density curve of the network, (b) matrix representation of the network, where the orange lines indicate the
borders separating different blocks. Here $\alpha=8$ and $\beta=0.6$.}
\label{fig8}
\end{figure}
\subsection{College football network}
The college football network describes games between Division I-A American college football teams in the year 2000~\cite{newman2004fast}. It has 115 nodes and 613 connections, leading to $\alpha=10$. From the region density curve in Fig.~9(a), we can conclude that the college football network only has community structure but has no CP structure.
By setting $\alpha=8$, the characteristics of region density curve is more clear and the peaks are easier to be observed (see Fig.~9(b)). For the football network, the inner connections in each community are not very dense, so is hard to achieve full connection in each community. For this purpose, we set $\beta=0.6$ to detect the community structure. According to our method, the network can be divided into 11 communities. The adjacency matrix of the network is shown in Fig.~9(c), which is very similar to the adjacency matrix shown in Fig.~1(a).
\begin{figure*}
\centerline{\includegraphics[width=6in]{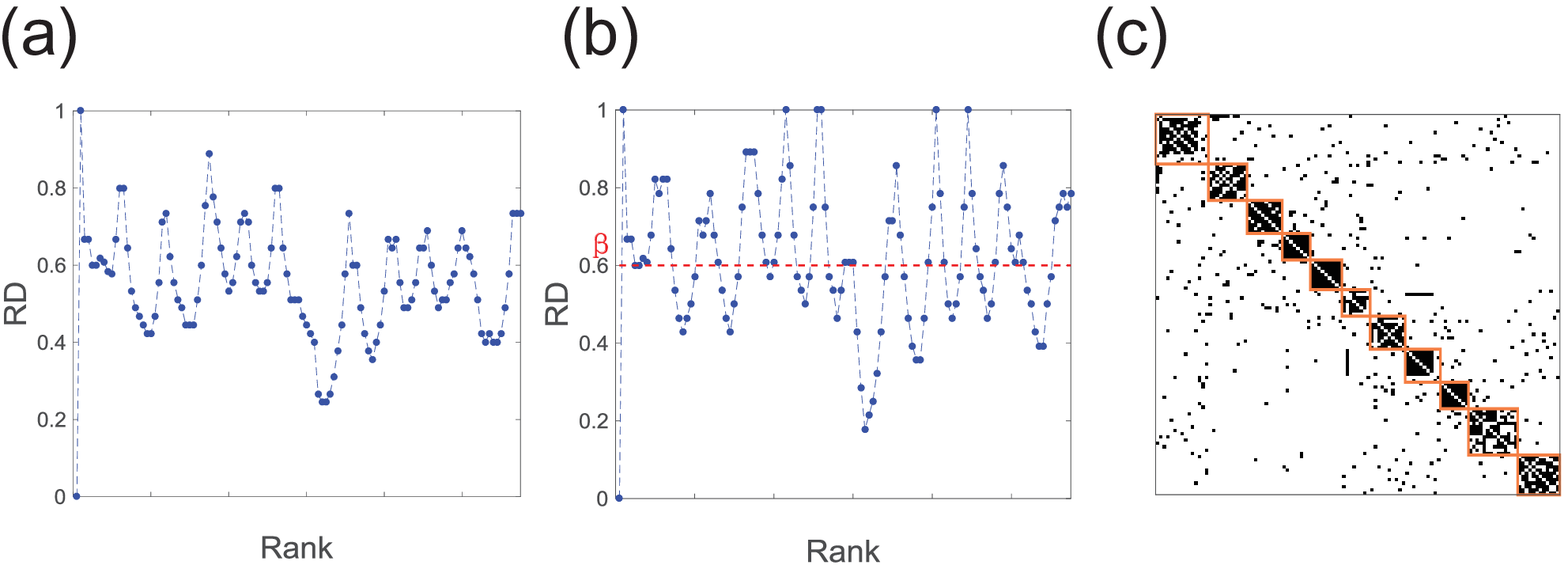}}
\caption{Detection on college football network. (a) region density curve of the network when $\alpha=10$, (b) region density curve of the network when $\alpha=8$, (c) matrix representation of the network when $\alpha=8$. Here $\beta=0.6$.}
\label{fig9}
\end{figure*}

\section{Comparison of different algorithms}

In this section, we compare the performance of our algorithm with other existing methods regarding the detection of CP structure and community structure, respectively. The normalized mutual information (NMI) index is used to measure the performance of different algorithms, which is defined as~\cite{pizzuti2012multiobjective}:
 \begin{eqnarray}\label{NMI}
NMI(A, B)=\frac{2I(A,B)}{H(A)+H(B)}.
\end{eqnarray}
Here $A$ and $B$ are the partition determined by algorithms and the real partition, respectively, $I(A,B)$ is the mutual information of $A$ and $B$. $H(A)$ and $H(B)$ are the entropy of $A$ and $B$, respectively. NMI is in the range of $[0,1]$ and equals to 1 only two partitions are totally coincident.

\subsection{Performance on detection of CP structure}

Since the study on the detection of CP structure has not been paid much attention, moreover, some existing algorithms were proposed to detect single CP structure and the number of core nodes should be given in advance. Thus, few algorithms can be used to fairly compare. Very recently, one algorithm aimed at detecting multiple pairs of CP structure was proposed by Sadamori Kojaku and Naoki Masuda (termed as KM algorithm)~\cite{kojaku2017finding}. The KM algorithm is to maximize one defined quality function. Moreover, in the KM algorithm, nodes were removed as residual nodes by considering the statistical significance of each CP pair. Here we do not check the statistical significance of the CP structure when we implement this algorithm, since the statistical significance is not considered in other algorithms. In the work, Kajuka \emph{et~al.} compare their algorithm with two other algorithms. One is BE-KL algorithm, which aims to detect a single CP structure by maximizing $Q_{BE}$ (a quality function based on the Pearson correlation coefficient to measure the similarity between the given partition and its ideal CP structure ) using the Kernighan-Lin algorithm~\cite{kernighan1970efficient}.  BE-KL algorithm mainly focuses on how to detect networks with single CP structure. The other algorithm is termed as Two-step algorithm, the network is first divided into non-overlapping communities by maximizing modularity using the Louvain algorithm~\cite{blondel2008fast}, then the core and periphery in each community are detected by BE-KL algorithm again. Here, we compare our algorithm with these three algorithms:  BE-KL, Two-step and KM algorithms on sets of synthetic networks, including networks with single CP structure and multiple pairs of CP structures, respectively.

\begin{figure}
\centerline{\includegraphics[width=3.5in]{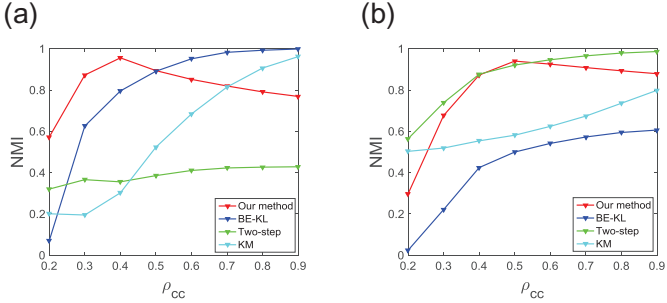}}
\caption{ NMI accuracy as a function of $P_{CC}$ on sets synthetic networks with different algorithms. (a) synthetic networks with single CP structure, (b) synthetic networks with two pairs of CP structures.}
\label{fig10}
\end{figure}

Since the ground truths about CP structures in real networks are not known, we generate some synthetic networks based on the stochastic block models~\cite{zhang2015identification,valles2016multilayer} to measure the performances of different algorithms. For sets of synthetic networks with single CP structure, the size of whole network, the size of core and the size of periphery are set as $N=600$, $N_c=100$ and $N_P=500$, respectively. For a network with typical CP structure, the connection probability among core nodes ($P_{CC}$) should be larger than or equal to the connection probability between core nodes and peripheral nodes ($P_{CP}$), and further larger than the connection probability among peripheral nodes ($P_{PP}$). Therefore, we let $P_{PP}=0.05$, and $P_{CP}=\frac{P_{CC}}{4}$, then sets of synthetic networks are generated by varying the value of $P_{CC}$. Two parameters in our method are set as $\alpha=\lfloor k\rfloor$ and $\beta=5\rho$, where $\rho=\frac{2M}{N(N-1)}$ with $M$ be the number of links. As shown in Fig.~\ref{fig10}(a), our method has better performance than the other algorithms when $P_{CC}$ is not very large, namely, the CP structure is not very strong. Otherwise, the performance of BE-KL algorithm is better than others when $P_{CC}$ is very large.

Several sets of synthetic networks with two pairs of CP structures are also generated to compare the performances of different algorithms. For each pair of CP structure, the parameters are set as: $N_c=50$, $N_P=250$, $P_{PP}=0.05$ and $P_{CP}=\frac{P_{CC}}{4}$, respectively. Moreover, the connection probability linking two pairs of CP structures is fixed as 0.01. Moreover, we set $\alpha=\lfloor k\rfloor$ and $\beta=10\rho$.  Fig.~\ref{fig10}(b) indicates that the Two-step algorithm gives rise to the best performance, and our method is better than the other two algorithms in most cases. For the Two-step algorithm, each step needs to maximize the corresponding quality function based on the global structure of network, which significantly increases the time complexity. Moreover, the performance of the Two-step algorithm on detection of network with single CP structure is very bad. The results on the synthetic networks with single CP structure and with multiple CP structures indicate that the performance of our method is generally better than other algorithms.

\subsection{Performance on detection of overlapping communities}
The performances of different algorithms on detecting overlapping communities are compared on three real networks and three LFR benchmark networks~\cite{lancichinetti2009community}( introduced in Sec.~\ref{synthetic}). The three real networks are the karate club network, dolphin social network and college football network, respectively. The ground truths about the community structure in these networks are known. The basic topological features of these networks are given in table~\ref{table1}.

\begin{table}[htbp]
\centering
\caption {The basic topological features of considered networks and the values of $\alpha$ and $\beta$ are chosen. $N$ and $M$ are the total numbers of nodes and links,
respectively. $NC$ is the number of communities. $\rho=\frac{2M}{N(N-1)}$ is the density of network. $\langle k\rangle$ is the average degree.}
\begin{tabular}{c|c|c|c|c|c|c|c}
\hline
Network& $N$ & $M$  &$NC$ & $\rho$ & $\langle k\rangle$&$\alpha$&$\beta$\\
\hline
Karate &34&78&2&0.1390&4.588&4&1 \\ \hline
Dolphins &62&159&2&0.0841&5.129&5&0.9 \\ \hline
Football &115&613&2&0.0935&10.048&10&0.6\\ \hline
Net1 &100&471&9&0.0952&9.42&9&0.6\\ \hline
Net2 &500&2402&10&0.0193&9.608&9&0.3\\ \hline
Net3 &1000&5001&10&0.01&10.002&10&0.3\\ \hline
\end{tabular}\label{table1}
\end{table}

We compare our method with four overlapping community detection algorithms: clique percolation
method (CPM) proposed by Palla \emph{et al.}~\cite{palla2005uncovering}, a local expansion and optimization algorithm (LFM) proposed by Lancichinetti \emph{et al.}\cite{lancichinetti2009community},  a fast overlapping community search (FOCS) method developed by Bandyopadhyay \emph{et al.}~\cite{bandyopadhyay2015focs}, and a fuzzy c-means clustering (FCM) algorithm proposed by Zhang \emph{et al}~\cite{Zhang2007Identification}. The comparisons of them are summarized in table~\ref{table2}. One can observe that the performance of our method on detecting overlapping communities generally outperforms other methods, especially on real networks.

\begin{table}[htbp]
\centering
\caption {The performances of different algorithms on detecting overlapping communities are compared on three real networks and three synthetic networks (Net1, Net2, Net3). The values of $\alpha$ and $\beta$ in our method are listed in the last two columns of table~\ref{table1}.}
\begin{tabular}{c|c|c|c|c|c}
\hline
Network &LFM&FCM&FOCS & CPM& Our method\\
\hline
Karate &0.6904& 0.6148 &0.2889& 0.2161& 0.9186 \\ \hline
Dolphins &0.7811& 0.8889& 0.2334& 0.1986 &0.8889 \\ \hline
Football &0.754&0.7357&0.6773&0.7471&0.7404\\ \hline
Net1 &0.9225&0.9624&0.8482&0.9846&0.9171\\ \hline
Net2 &0.7008&0.9535&0.0248&0&0.5011\\ \hline
Net3 &0.5731&0.9576&0&0&0.4536\\ \hline
\end{tabular}\label{table2}
\end{table}

\section{Conclusions}
\label{sec:con}
In this work, we have proposed a unified method to detect CP structure and community structure in networks.
This method is effective not only on  single CP structure, but also on community structure and multiple pairs of CP structures,
and further on finding active nodes and overlapping nodes in networks.
Also, the role of each node is assigned.
In addition, our method does not need to fix the size of core in advance, where some sequential cores can form a larger core.
Therefore, our method provides a tool for the identification of meso-scale structures in network, and may also provide some inspirations in identifying the influential nodes.

Of course, many places are worthy of
future study. On the one hand, we set $\alpha=\lfloor \langle k\rangle\rfloor$ and $\beta=1$ in most cases. For each network, how to choose the values of $\alpha$ and $\beta$ is a non-negligible problem. Of course, as we have discussed, many rich structures can be observed if the values of $\alpha$ and $\beta$ are properly chosen. On the other hand, previous results have demonstrated that the network structure has fundamental effect on its collective dynamics, for instance, networks with scale-free structure can promote the outbreaks of infectious diseases~\cite{gang2005epidemic,pastor2015epidemic} and reduce the transportation efficiency~\cite{yan2006efficient}. The community structure also significantly affects its collective dynamics, such as synchronization~\cite{zhou2007phase}, spreading of epidemic~\cite{liu2005epidemic}, evolutionary games~\cite{chen2007prisoner}, and so forth. As a result, how the CP structure affects the collective dynamics is an important issue that deserve in-depth studies. For example, Sim\'{o}m \emph{et~al.} have found that the CP structure can induce double percolation phase transition under certain condition~\cite{colomer2014double}.

\section{Appendix}
\begin{pseudocode}[plain]{GeneralFramework}{G(V, E),\beta}
 \COMMENT{\mbox{$G(V, E)$ is a network , $\beta$ is a threshold  }}\\
 U\GETS \CALL{RerankNodes}{G}\\
 \alpha \GETS \lfloor <k>\rfloor\\
 \FOREACH \ u_{i} \in U \DO

 \BEGIN

 \CALL{RD}{u_{i}} \GETS \mbox{Compute region density of node $u_i$}\\
                         \mbox{ by Eq.~(\ref{eq3}) } \\
 \END\\
Cset\GETS \CALL{FindCoreSet}{G,U,\beta}\\

CPset\GETS \CALL{FindCPSet}{G,Cset,NumC}\\
\RETURN{Cset,CPset}\label{algorithm1}
\end{pseudocode}

\begin{pseudocode}[plain]{RerankNodes}{G(V, E)}
 \COMMENT{$G(V, E)$ is a network}\\

 U\GETS \emptyset , V'\GETS V\\
 \COMMENT{Select the node with the highest closeness}\\
 \mbox{centrality as the first node.}\\

 \mbox{$u\leftarrow $max$(V.closeness)$} \\
 \mbox{Add $u$ into $U$ }   \\

 \mbox{Remove $u$ from $V'$} \\

 \COMMENT{Select the node with the most connections }\\
 \mbox{with the nodes in $U$, if more than one node were found,}\\
 \mbox{ we choose the node with the maximum degree.} \\

\WHILE V' \neq \emptyset  \DO
\BEGIN
     \FOREACH  i \in neighbor\_U \DO

     \BEGIN
          P(i)\GETS \mbox{Compute priority of node $i$ }\\
                                 \mbox{ by Eq.~(\ref{eq1}) } \\
     \END\\
     \mbox{$u \leftarrow \mathop{\max}\limits_{i}(P(i))$}\\

     \mbox{Add $u$ into $U$}\\

     \mbox{Remove $u$ from $V'$}\\

\END\\

\RETURN{U}\label{algorithm2}
\end{pseudocode}

\begin{pseudocode}[plain]{FindCoreSet}{G(V, E),U,\beta}
 \COMMENT{$G(V, E)$ is a network,U is a set of node, $\beta$ is}\\
 \mbox{a threshold} \\

 \mbox{$Cset \leftarrow \emptyset$, $NumC \leftarrow 1$}\\

 \COMMENT{if the values of RD for two sequential nodes  }\\
 \mbox{are greater than or equal to $\beta$, the two cores are merged }\\
 \mbox{ as a single one.}\\

\FOR i \GETS \alpha \TO numNode \DO
\BEGIN
   \IF RD(u_{i}) \geq \beta \THEN
       \BEGIN
          \mbox{$Cset(NumC)\leftarrow Cset(NumC)\cup$}\\
          \mbox{$\left\{u_{i-\alpha+1},...,u_{i}\right\}$}\\
          \ELSE
              \BEGIN
                  \IF RD(u_{i-1}) \geq \beta \ and \  i >\alpha \THEN
                  \mbox{$NumC\leftarrow NumC+1$}\\
              \END\\
       \END\\

\END\\

\IF RD(u_{numNode})<\beta \THEN
   \mbox{$NumC\leftarrow NumC-1$}\\
\RETURN{Cset,NumC}\label{algorithm3}
\end{pseudocode}

\newpage
\begin{pseudocode}[plain]{FindCPSet}{G(V, E),Cset,NumC}
 \COMMENT{$G(V, E)$ is a network, Cset is a set of core }\\

\mbox{$CPset\leftarrow Cset$}\\

\mbox{$class\leftarrow 0$}\\

\mbox{$nodeCset.class\leftarrow class$}\\

\mbox{$NodeP\leftarrow V-nodeCset$}\\
\COMMENT{Allocate the periphery nodes}\\

\WHILE NodeP \neq \emptyset  \DO
\BEGIN
     \mbox{$neighbor\_CPset.class\leftarrow class+1$}\\

     \FOREACH  j \in neighbor\_CPset \DO

     \BEGIN
         \IF connections(j,CPset(i))\  is\ max \THEN
          \mbox{Add $j$ into $CPset(i)$}\\
     \END\\
     \mbox{Remove $neighbor\_CPset$ from $NodeP$}\\

\END\\

 \mbox{$NodeActive\leftarrow $$node$ $in$ $different$ $CPset(i)$} \\

 \COMMENT{Re-distribute active nodes}\\

 \mbox{Remove $NodeActive$ from $CPset(i)$}   \\

\FOREACH  j \in NodeActive \DO
\BEGIN
     \IF connections(j,CPset(i))\ is\ max \THEN
     \mbox{Add $j$ into $CPset(i)$}\\
\END\\

\RETURN{CPset}\label{alogrithm4}
\end{pseudocode}

\section*{Acknowledgments}

This work is funded by the NSFC
(Grant Nos. 61473001, 11331009, 61672033), and partially supported by the Young Talent Funding of Anhui Provincial Universities (gxyqZD2017003). BBX is also supported by two research projects from Anhui University (Grant Nos. Y01002451, Y01002430).

%

\end{document}